\documentclass[reprint, amsmath,amssymb, aps, nofootinbib]{revtex4-2}

\usepackage{graphicx}
\graphicspath{{./}}
\usepackage{dcolumn}
\usepackage{bm}
\usepackage{tikz,xcolor,hyperref}

\definecolor{lime}{HTML}{A6CE39}
\DeclareRobustCommand{\orcidicon}{
	\begin{tikzpicture}
	\draw[lime, fill=lime] (0,0) 
	circle [radius=0.16] 
	node[white] {{\fontfamily{qag}\selectfont \tiny ID}};
	\draw[white, fill=white] (-0.0625,0.095) 
	circle [radius=0.007];
	\end{tikzpicture}
	\hspace{-2mm}
}
	
\foreach \x in {A, ..., Z}{\expandafter\xdef\csname orcid\x\endcsname{\noexpand\href{https://orcid.org/\csname orcidauthor\x\endcsname}
			{\noexpand\orcidicon}}
}

\begin{document}

\preprint{APS/123-QED}

\title{Forced and natural dynamics of a clamped flexible fiber in wall turbulence}

\author{Giulio Foggi Rota \orcidA{}}
\author{Morie Koseki \orcidB{}}
\author{Riya Agrawal \orcidC{}}
\altaffiliation[Currently at ]{United College of Engineering and Research, Allahabad, 211010, Uttar Pradesh, India.}
\author{Stefano Olivieri \orcidD{}}
\altaffiliation[Currently at ]{Universidad Carlos III de Madrid, Leganés, 28911, Madrid, Spain.}
\author{Marco Edoardo Rosti \orcidE{}}
\email[E-mail for correspondence: ]{marco.rosti@oist.jp}
\affiliation{Complex Fluids and Flows Unit, Okinawa Institute of Science and Technology Graduate University, 1919-1 Tancha, Onna-son, Okinawa 904-0495, Japan.}

\begin{abstract}

We characterize the dynamical behavior of a clamped flexible fiber immersed in wall turbulence over a wide range of natural frequencies ($f_{nat}$) by means of direct numerical simulations.
Only two flapping states are possible: one where the fiber oscillates at the characteristic frequency of the largest turbulent eddies ($f_{turb}$) and another where the natural structural response dominates. The former is obtained in the more flexible cases ($f_{nat}<f_{turb}$), while the latter in the more rigid ones ($f_{nat}>f_{turb}$).
We observe here for the first time that in the turbulence dominated regime the fiber always sways at a frequency proportional to the largest scale of the flow, regardless of its structural parameters. 
The hindrance of the clamp to the wall prevents the synchronization of the fiber with turbulent eddies of comparable size.

\end{abstract}

\maketitle


Atmospheric winds and marine currents often interact with the solid surface of the Earth, giving rise to geophysical boundary layers.
Those are frequently disturbed by the presence of slender and flexible obstacles, either natural or artificial, which protrude from the ground. 
The gentle swaying of a tree in the wind, or the dangerous oscillations of a pylon, are only few demonstrations of the rich dynamical behavior which emerges as a consequence of the complex fluid-structure interaction discussed in this Letter.

When a multitude of slender objects is uniformly distributed over a region which is wide compared to their height (e.g., trees in a forest or seaweed on the seafloor, but also cilia on a membrane) a canopy is attained \cite{raupach-thom-1981, finnigan-2000, nepf-2012a, brunet-2020, loiseau-etal-2020}.
With some noticeable exceptions in the field of microfluidics \cite{wang-etal-2022-2}, the flow above and within a canopy is typically turbulent.
The turbulent flow established within and immediately above a canopy differs significantly from a conventional boundary layer, since the structure of turbulence is altered \cite{poggi-etal-2004, ghisalberti-nepf-2004, sharma-garciamayoral-2020, olivieri-etal-2020-2, monti-etal-2022, nicholas-omidyeganeh-pinelli-2022} and the diffusion of passive species (like suspended sediments, seeds and pollutants) is enhanced \cite{luhar-rominger-nepf-2008, mossa-etal-2017, makar-etal-2017, shnapp-etal-2020, quin-etal-2022}.
Recently, considerable effort has been spent on investigating the features of the flow and the dynamics of the flexible canopy elements \cite{tschingale-etal-2021, wang-etal-2022-1, he-liu-shen-2022, wang-etal-2022-1,monti-olivieri-rosti-2023}; nevertheless, the dynamics of a single constitutive element has not been characterised in detail yet.
This study therefore represents a step back from the existing body of works and aims at characterizing numerically the behavior of a clamped flexible fiber immersed in wall turbulence for different values of its rigidity, length and density. 
We provide solid grounds for a thorough understanding of more complex systems composed by several elements, where the relation between their individual and collective dynamics becomes important.

\begin{figure}[h!]
\centering
\includegraphics[scale=0.08]{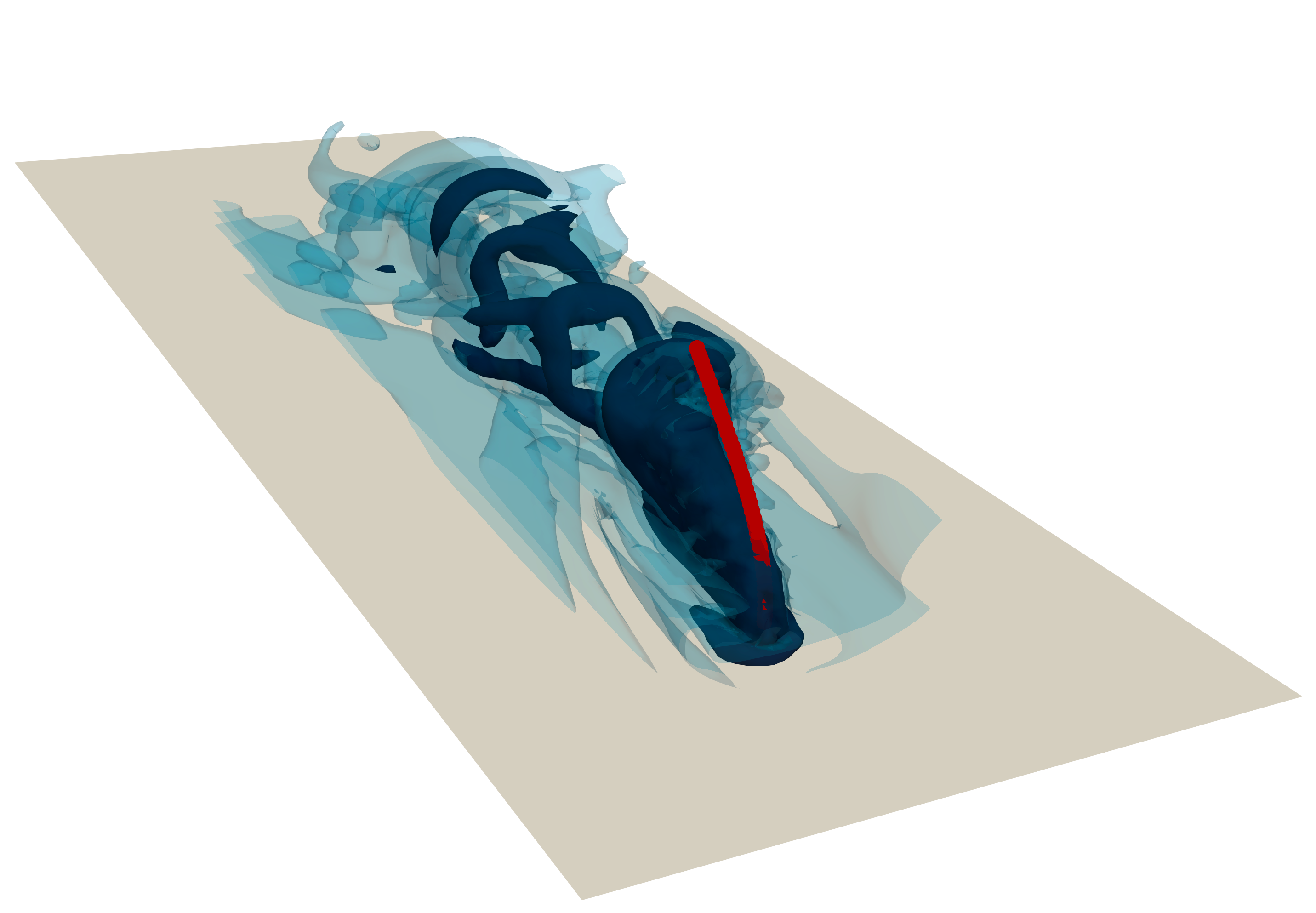}
\caption{Flexible fiber clamped in a turbulent wall flow. The fiber is colored in red, while the flow structures behind that are visualized as blue iso-surfaces of the vorticity magnitude, $|\mathbf{\omega}|$. Light and dark correspond to $|\mathbf{\omega}|=6.5 U_b/H$ and $|\mathbf{\omega}|=15.5 U_b/H$, respectively.}
\label{fig:cover}
\end{figure}

In this Letter we show how the dynamical response of a flexible fiber immersed in a \textit{wall bounded turbulent flow} (like the one depicted in Fig. \ref{fig:cover}) can only vary among two different regimes, depending on its structural properties.
Specifically, the fiber is observed to either oscillate at its natural frequency, $f_{nat}$, or sway at a rate comparable to that of the largest turbulent eddies in the flow, $f_{turb}$.
While the former scenario is attained when $f_{nat}/f_{turb}\gg1$, the latter corresponds to a condition where $f_{nat}/f_{turb}\ll1$ and the flapping frequency of the fiber becomes independent from its structural properties.

To tackle this fluid-structure interaction problem we solve numerically a modified version of the Euler-Bernoulli equations, which well describe the dynamics of an inextensible elastic fiber, coupled to the Navier-Stokes equations, which fully represent the motion of the incompressible Newtonian fluid in which the isolated fiber is immersed. 
The setup adopted is that of a conventional turbulent channel flow at a Reynolds number $Re=U_b 2H/\nu=5600$, where a fluid of kinematic viscosity $\nu$ streams with a mean bulk-averaged velocity $U_b$ along the positive $x$-direction, between two no-slip planes at a distance $2H$ in the $y$-direction.
We investigate the behavior of almost rigid fibers as well as flexible ones, spanning different values of the Cauchy number, $Ca$, defined as the ratio among the force exerted by the fluid and the restoring elastic force opposed by the fiber, $Ca=(\rho_f d h^3 U_b^2)/(2\gamma)$, where $\rho_{f}$ is the volumetric density of the fluid, $d$ is the cross-section diameter of the fiber, $h$ is its length, and $\gamma$ its bending rigidity.
In particular, we vary the Cauchy number by varying the length of the fiber, $h$, or its bending rigidity, $\gamma$, and the density ratio between fiber and fluid, $\rho_s/\rho_f$, by varying the fiber density, $\rho_s$. 
Further details on the values of the relevant parameters employed throughout our simulations and about our numerical setup are provided in the Supplementary Material~\footnote{Technical details on the numerical methods and the setup of our simulations, along with additional results, are provided in the Supplementary Materials, with references \cite{rosti-etal-2018, mazzino-rosti-2021,kim-moin-1985,dorr-1970,yu-2005,banaei-rosti-brandt-2020,huang-etal-2007,peskin-2002,monti-olivieri-rosti-2023,yoshihiko-etal-1991,jimenez-moin-1991,kim-moin-moser-1987,jin-ji-chamorro-2016}.}.

The most evident effect of the flow on a clamped fiber is the deflection of its time-averaged position: the fiber reconfigures its shape bending in the streamwise direction proportionally to the value of $Ca$ and experiences a reduced drag compared to its typical quadratic scaling with the velocity, especially when the hydrodynamic force is most intense \cite{alben-shelley-zhang-2002, gosselin-langre-machadoalmeida-2010, luhar-nepf-2011}.
\begin{figure}
\centering
\includegraphics[scale=0.45]{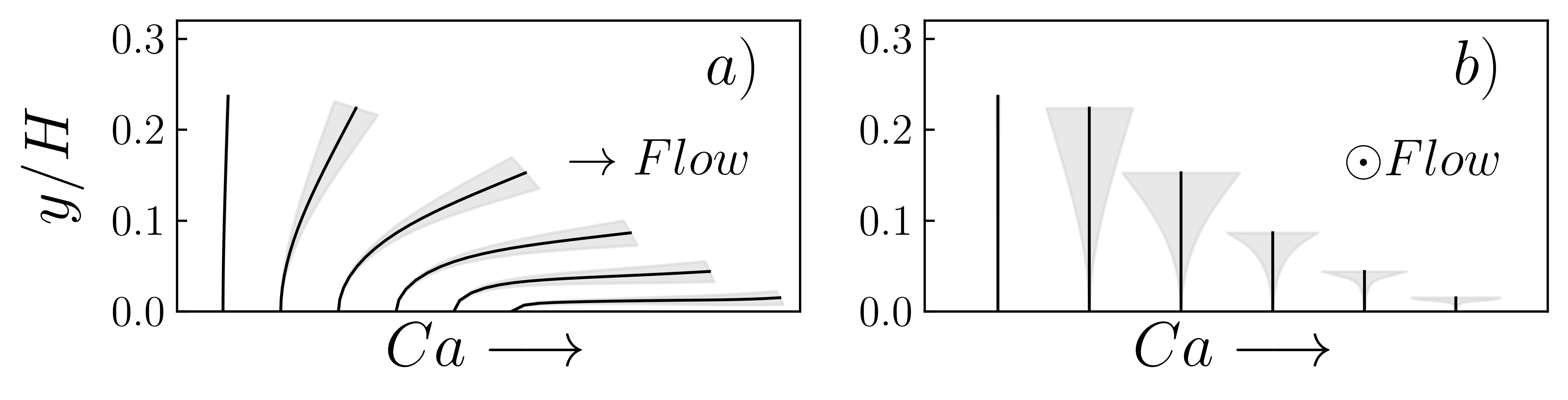}
\caption{Time-averaged configuration (black lines) of the fiber $0.25H$ long for logarithmically increasing values of $Ca$ in the range $[3.1 \cdot 10^{-2}, 3.1 \cdot 10^{3}]$. Panel $a$/$b$ shows the side/front-view. The fiber is deflected by the mean flow according to its rigidity, hence oscillates about its time-averaged configuration. Grey regions show the root-mean-square of the displacement from the time-averaged configuration.}
\label{fig:fibers}
\end{figure}
The fiber reconfiguration is shown in Fig.~\ref{fig:fibers}, where black lines denote its time-averaged positions for different values of $Ca$.
Instead, the grey regions represent the envelope in which the fiber oscillates.
The amplitude of the oscillations increases while moving away from the root and depends on both the structural properties of the fiber (namely, $Ca$ and $\rho_s$; more details on this in the Supplementary Material) and the turbulent state of the flow at the wall-normal distance where the fiber is located, overall exhibiting a non-monotonic trend.

To better understand the relation between the motion of the fluid and that of the fiber, we compare the probability density functions ($PDFs$) \footnote{The $PDF$ of a random variable $x$ at a given point (or sample) provides the relative likelihood of $x$ being equal to the sample. It is a well defined mathematical tool to describe random processes, like the ones observed in this Letter.} of their respective velocities.
For the fiber, we consider the velocity of the tip (where the motion is most pronounced) and compute the \textit{Lagrangian}-$PDF$ from its time history.
For the fluid, instead, we take into account the velocity fluctuations in a channel at the same $Re$ without the fiber and compute the \textit{Eulerian}-$PDF$ from the space and time signal measured at a wall-normal position correspondent to that of the deflected tip. 
On comparing the Eulerian and Lagrangian $PDFs$, two different behaviors are observed when varying the rigidity of the fiber.
\begin{figure}
\centering
\includegraphics[scale=0.45]{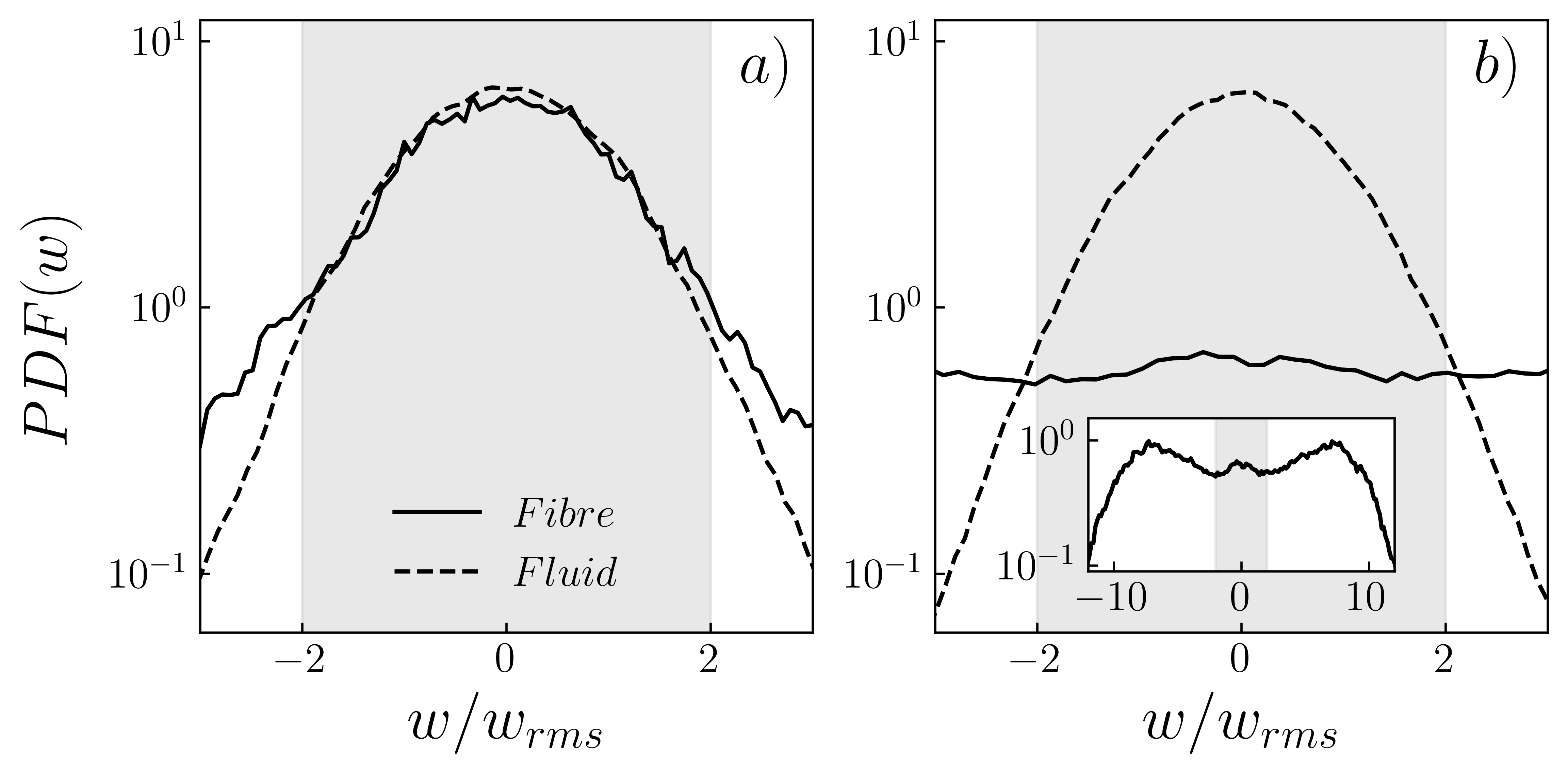}
\caption{Different regimes of motion of the fiber, identified with the $PDFs$ of the spanwise velocity component $w$ (i.e., the only one not directly affected by the presence of the mean flow and the inextensibility constraint, or by the confinement of the walls). 
Solid lines refer to the data taken from the fiber tip velocity, while dashed lines are associated to the flow velocity fluctuations measured at a wall distance corresponding to the mean position of the deflected fiber tip. 
The $rms$ of $w$ at such position, denoted $w_{rms}$, is adopted to adimensionalize $w$ itself. Panel $a$ refers to a flexible fiber ($Ca = 3.1 \cdot 10^{2}$, $h=0.25H$, $\rho_s=1.08\rho_f$) for which the two $PDFs$ overlap, hence suggesting that the fiber is moving with the flow. Panel $b$, instead, refers to a more rigid fiber ($Ca = 3.1$, $h=0.25H$, $\rho_s=1.08\rho_f$) for which the two $PDFs$ are radically different, thus hinting to an independent motion of the fiber. On a wider scale (inset), the $PDF$ of the fiber appears bimodal and compatible with nearly-sinusoidal oscillations.}
\label{fig:PDFs}
\end{figure}
At a high value of $Ca$, the $PDFs$ of the fiber overlap well with those of the flow up to absolute values of the velocity twice as big as the root-mean-square ($rms$) of the flow velocity fluctuations for both the spanwise (panel $a$ of Fig.~\ref{fig:PDFs}) and wall-normal components.
This is indicative of a regime where the fiber, deflected forward by the current, sways coherently with the turbulent fluctuations in the directions not forced by the mean flow and does not exhibit any independent dynamics.
The flow also dominates the streamwise dynamics of the fiber, but the inextensibility constraint and the clamp to the wall hinder its motion. 
After reaching a critical value of $Ca$ (which lays roughly in the middle of our investigated range) all the $PDFs$ of the fiber start widening and drift away from those of the flow (panel $b$ of Fig.~\ref{fig:PDFs}), as the fiber moves independently.
Further decreasing $Ca$ always leads to states where the $PDFs$ of the flow and of the fiber do not match.

\begin{figure}
\centering
\includegraphics[scale=0.43]{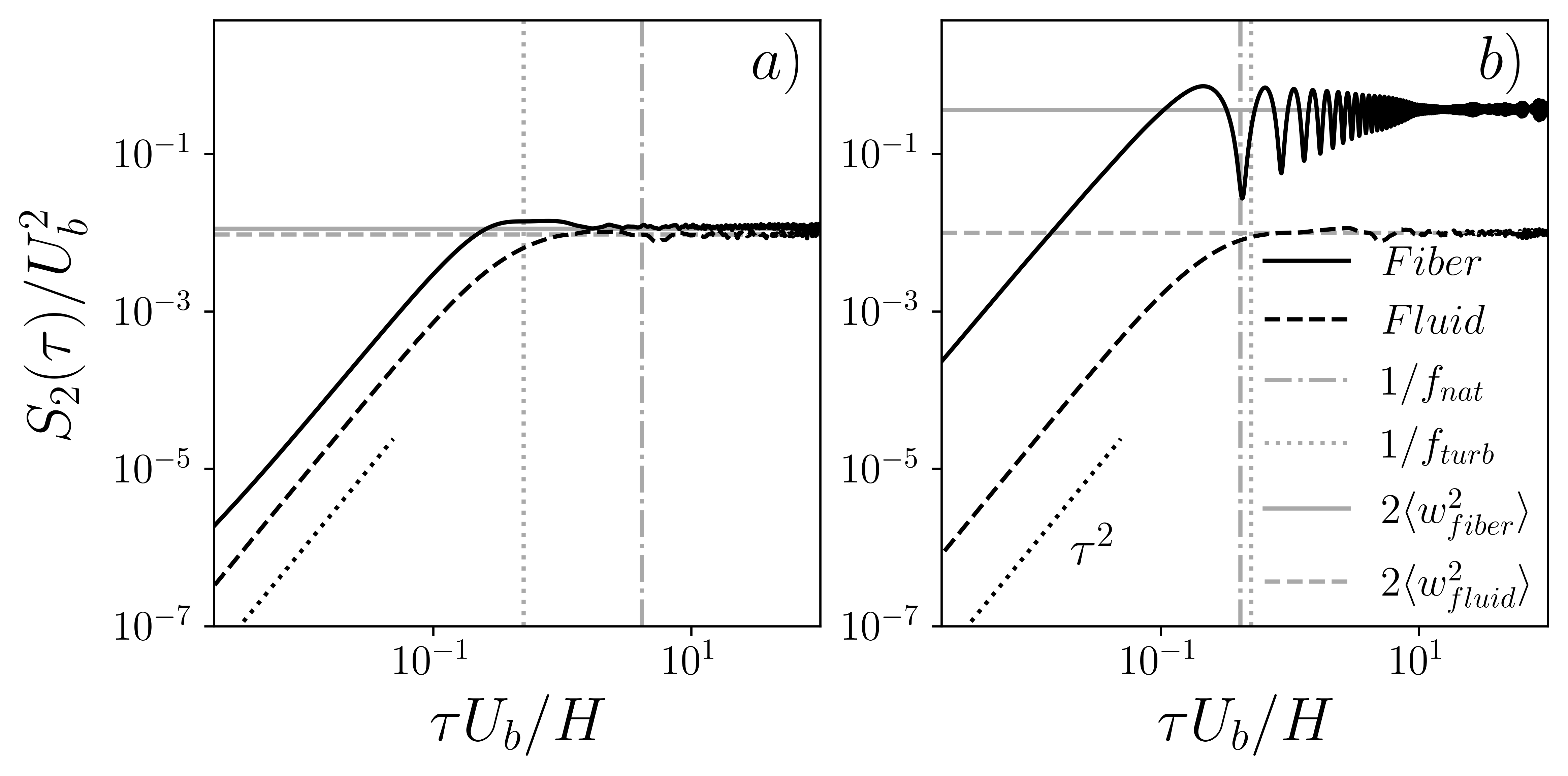}
\caption{Second order structure function of the spanwise velocity component $w$. Solid black lines refer to the Lagrangian data taken from the fiber tip velocity, while dashed black lines are associated to the Eulerian flow velocity fluctuations measured at a wall distance corresponding to the mean position of the deflected fiber tip. Panel $a$ refers to a flexible fiber ($Ca=3.1\cdot 10^{2}$, $h=0.25H$, $\rho_s=1.08\rho_f$) for which $S_2$ saturates at $\tau \approx 1/f_{turb}$ reaching a value $2\langle w_{fiber}^2 \rangle \approx 2\langle w_{fluid}^2 \rangle$, highlighting the passive motion of the fiber in the flow. Panel $b$, instead, refers to a more rigid fiber ($Ca=3.1$, $h=0.25H$, $\rho_s=1.08\rho_f$) for which the natural response is visible and the saturation value does not correspond to that of the fluid, suggesting an independent motion of the fiber.}
\label{fig:structFun}
\end{figure}

The existence of two different regimes of motion for the fiber, as well as their main features, can also be appreciated comparing the second order Lagrangian temporal structure function of the fiber tip spanwise velocity to the Eulerian one of the fluid, at the mean position of the deflected fiber tip.
In Fig.~\ref{fig:structFun} we therefore observe $S_2(\tau)=\langle (w(t+\tau)-w(t))^2 \rangle$ for the same two fibers considered in Fig.~\ref{fig:PDFs}. 
In both cases, for the smallest values of the separation $\tau$, $S_2 \sim \tau^2$ follows directly from the Taylor expansion of $S_2$ as $\tau$ goes to $0$: the fluctuating signal $w$ is smooth over those time scales.
The signal decorrelates ($\langle w(t+\tau) - w(t) \rangle \approx 0$) for larger values of $\tau$ and the behaviour of the fiber emerges.
In the most flexible case (panel $a$), the Lagrangian structure function associated to the motion of the fiber saturates at $\tau \approx 1/f_{turb}$ reaching a value of $2\langle w_{fiber}^2 \rangle \approx 2\langle w_{fluid}^2 \rangle$, suggesting that the motion of the fiber is coupled to that of the fluid.
In the most rigid case (panel $b$), instead, the fiber exhibits its natural response, which is decoupled from the turbulent fluctuations of the flow and characterised by nearly-sinusoidal oscillations (as confirmed by the bimodal nature of the fiber PDF on the larger scale reported in the inset of Fig.~\ref{fig:PDFs}$b$).
The Lagrangian structure function associated to the motion of the fiber has local minima at $\tau \approx 1/f_{nat}$ and higher harmonics (where $f_{nat} \approx 3.516/(d h^2) \sqrt{\gamma/(\rho_{s}\pi^3)}$ is the natural frequency of the fiber), and saturates at a different value from that of the fluid.

Interestingly, our results bear similarities with previous investigations in the field of vortex-induced vibration (VIV) \cite{bearman-1984}.
Despite the significant differences in the setup (our clamped fiber is extremely slender and flexible, while VIV investigations mainly deal with more rigid and intrusive bluff bodies, with usually only one or two degrees of freedom), we identify two regimes of motion: one in which the fiber follows the flow and one in which it independently oscillates at its natural frequency.
We are now keen to better investigate and understand the dynamics of the fiber by sampling the motion of its tip for a sufficiently long time span, removing any initial transient and looking at the frequency spectra.
After inspecting the Fourier's transforms of all the displacement and velocity components, we decide to focus once again on the spanwise velocity: conclusions similar to those drawn in the following could also be reached by observing the other components, while dealing with a more disturbed signal affected by external constraints.
\begin{figure}
\centering
\includegraphics[scale=0.43]{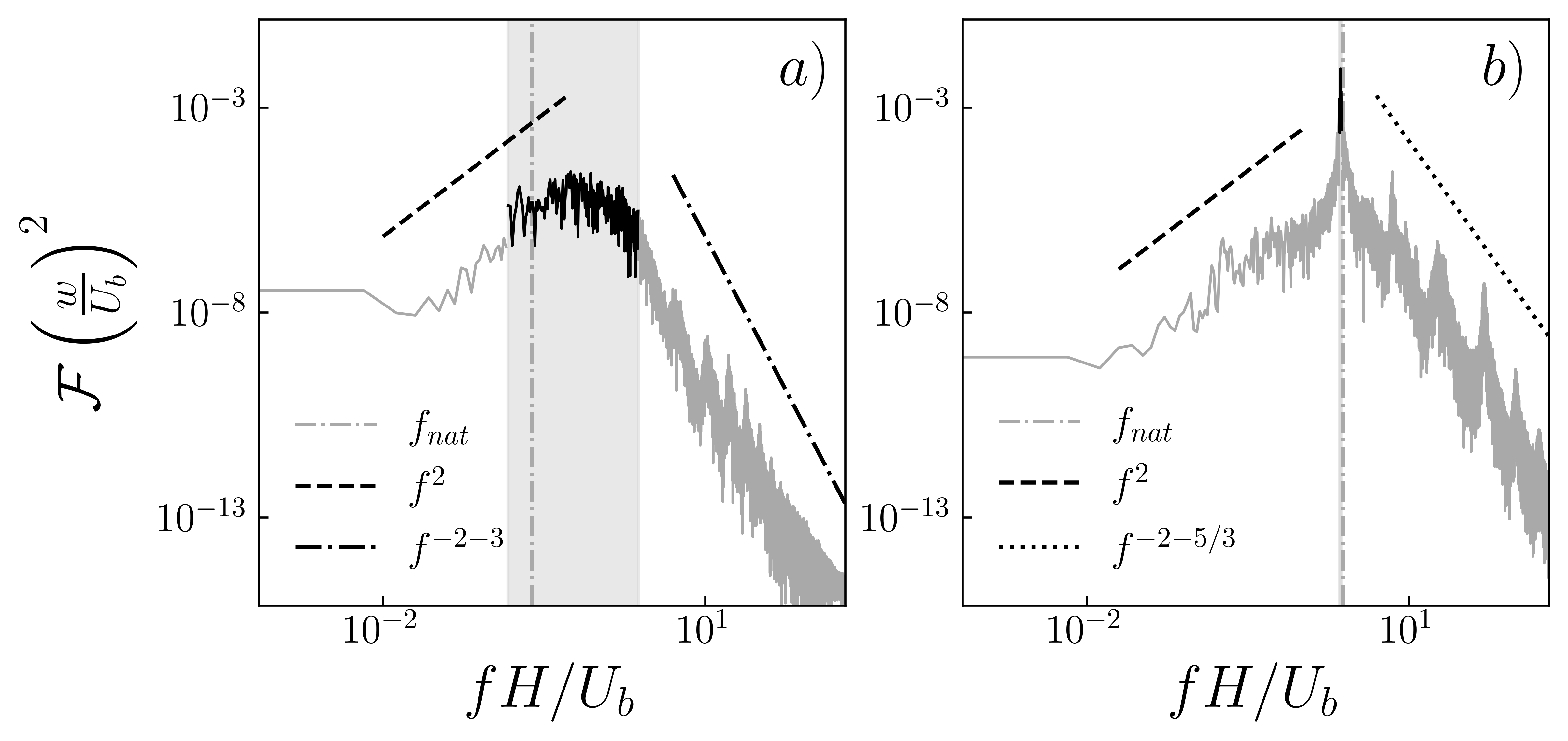}
\caption{Spectra of the tip spanwise velocity $w$ for $a$) a flexible ($Ca = 3.1 \cdot 10^{2}$, $h=0.25H$, $\rho_s=1.08\rho_f$) and $b$) a more rigid ($Ca = 3.1$, $h=0.25H$, $\rho_s=1.08\rho_f$) fiber. In the former, the broad-band peak of the signal spans roughly a decade of frequencies, while in the latter the peak is narrow and more prominent; we emphasize this difference shading the region among the two frequencies at which the signal achieves for the first time a value one decade below its maximum, approaching the peak from the left and from the right. By representing the natural frequency of the fiber with a vertical dashed line, we confirm that the more rigid fiber (in panel $b$) is exhibiting its natural response. Both spectra follow a $f^2$ scaling in the low frequency region by construction. After the peak, instead, the fiber exhibits a $f^{-2+\xi}$ decay, with $\xi=-3$ for the flexible fiber, characteristic of a smooth regime, and with $\xi=-5/3$ for the more rigid one, typical of fully developed turbulence.}
\label{fig:spectra}
\end{figure}
Two scenarios are possible, as shown in Fig.~\ref{fig:spectra}: when the fiber is flexible and sways with the flow (as previously established from the $PDFs$) the spectrum exhibits a broad-band peak (panel $a$), spread over roughly a decade.
Instead, when the fiber is more rigid and sways at its natural frequency (as previously established from the second order structure functions), the peak is narrower and more prominent (panel $b$).
In this second case, we confirm that the frequency of the peak corresponds to the natural frequency of the fiber.

The spectra always follow a $f^2$ scaling in the low frequency region by construction, being the transform of a Lagrangian velocity signal, while at high-frequencies they decay according to a $f^{-2+\xi}$ law \cite{jin-ji-chamorro-2016}~\footnote{The theory of \citet{jin-ji-chamorro-2016}, well supported by experimental evidence, provides the general model of the transfer function associated to the velocity oscillations of the structure adopted in this study.}.
When the fiber is more flexible, it is deflected in the lower part of the viscous wall region ($0.05 \lesssim y/H \lesssim 0.1$) by the mean flow and oscillates under the forcing of a smooth turbulent field, thus hinting towards a $\xi = -3$ decay of the spectrum \cite{frisch-1995}.
This argument is well supported by the high-frequency range of our data, in panel $a$ of Fig.~\ref{fig:spectra}.
When the fiber is more rigid, instead, it spans the buffer layer and turbulence takes over, thus justifying the recovery of the conventional $\xi = -5/3$ exponent.
We therefore point out the interesting counterposition between the dynamics of a more rigid fiber, dominated by its natural response but ``reflecting" the turbulent flow in the high frequency range, and that of a more flexible one, dominated by the flow but forced by the smooth turbulent field to which the fiber is exposed as a consequence of its deflection in the near wall region.
The temporal spectra of the turbulent kinetic energy at the mean position of the deflected fiber tip, representing the forcing to which the fiber is exposed, are reported in the Supplementary Material.

\begin{figure}
\centering
\includegraphics[scale=0.5]{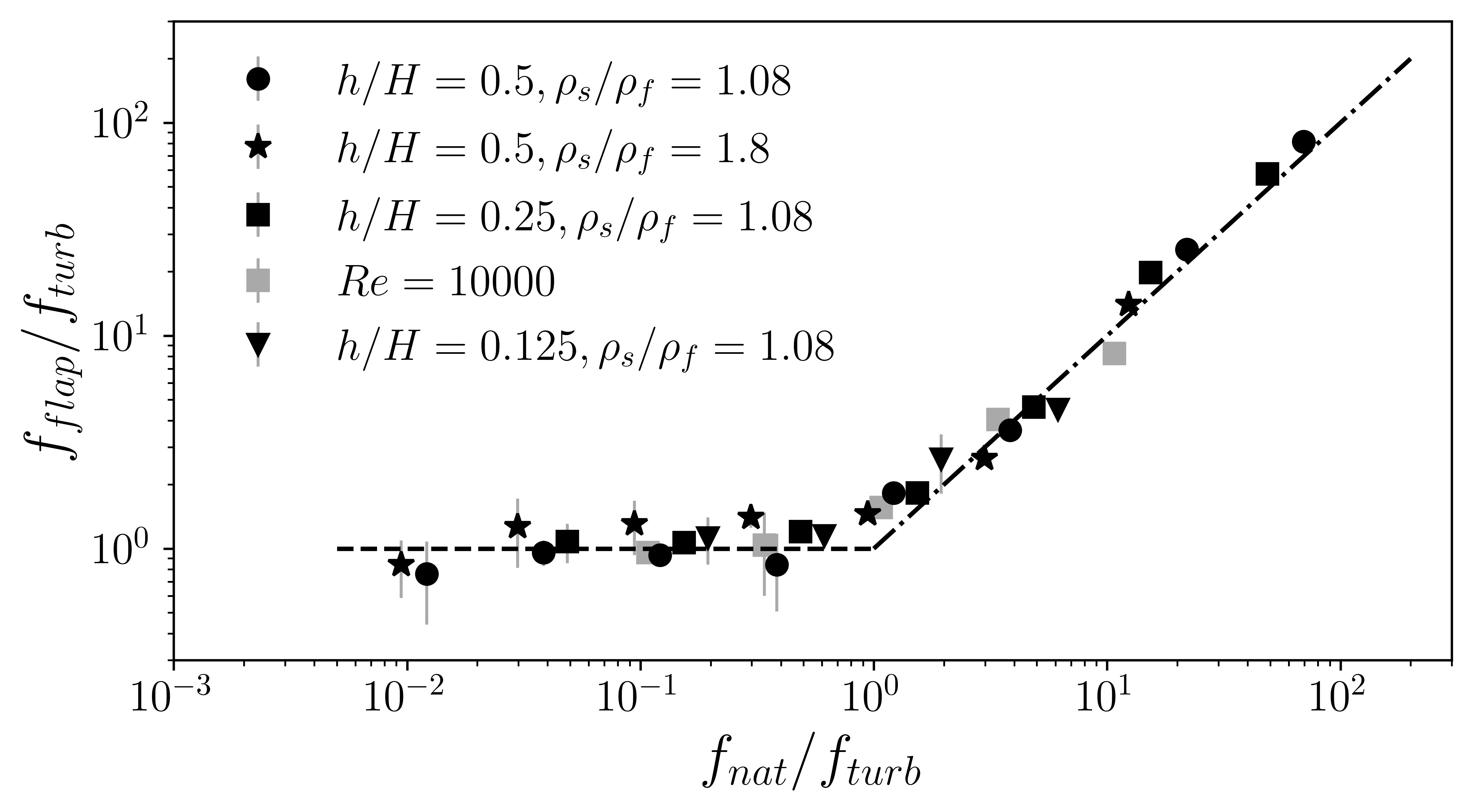}
\caption{Values of the spanwise flapping frequency ($f_{flap}$) as a function of the fiber natural frequency ($f_{nat}$). 
All frequencies are scaled with respect to the turbulent frequency measured by the most flexible filaments,  $f_{turb}= 0.5 {U_b}/{H}$.
The error bars represent the variation in the measured value of $f_{flap}$ when halving the time history.}
\label{fig:freqCollapse}
\end{figure}
The analysis of the single spectrum conducted above proves informative but not exhaustive: the location of the transition between the two oscillation regimes remains unknown and the physical meaning associated to the frequency of the spectral peak $f_{flap}$ is unclear in the more flexible cases.
To tackle these issues, we investigate multiple values of the parameters $Ca$, $h/H$ and $\rho_s/\rho_f$; we therefore extract $f_{flap}$ for each case \footnote{All the  values of $f_{flap}$ are extracted as the maximum location of the logarithm of the Fourier's spectrum of $w^2$, where only peaks with a prominence of $0.5$ computed using a smoothing kernel of $10$ points are compared to extract the highest. We have assessed the robustness of the results from the choice of such parameters.} and compare the outcomes. For the highest values of $Ca$ considered in this study, the fiber is observed to oscillate at $f_{flap} \approx 0.5 {U_b}/{H}$ regardless of its structural parameters.
We therefore set $f_{turb}= 0.5 {U_b}/{H}$ and map every case on a $f_{flap}/f_{turb}$ vs $f_{nat}/f_{turb}$ diagram in Fig.~\ref{fig:freqCollapse}, where $f_{nat}$ is a direct consequence of the choice of the structural parameters and it scales as ${f_{nat}}/{f_{turb}} \propto Ca^{-0.5}$ upon fixing $h/d$ and $\rho_s/\rho_f$.
As shown in Fig.~\ref{fig:freqCollapse}, the two regimes of motion of the fiber appear as two well distinct branches on the diagram: for $f_{nat}/f_{turb}\ll1$ the data plateaus, confirming our previous observation that $f_{flap}\cong f_{turb}$, while for $f_{nat}/f_{turb}\gg1$ the data collapse on a line with unitary slope, highlighting that $f_{flap}\cong{f_{nat}}$.
The horizontal branch therefore corresponds to the condition in which the fiber is compliant with the turbulent fluctuations of the flow: independently from the choice of the structural parameters, the dominant spectral component of the tip motion lays at a frequency dictated by turbulence.
The apparent absence of any relation among the value of $f_{turb}$ sampled by the fiber and its structural characteristics is a novel phenomenon, demanding for further experimental verifications.
On the other hand, the structural parameters play a major role along the ascending branch of the diagram, where they set the value of the natural frequency at which the fiber oscillates. 
The transition between the two regimes occurs at $f_{nat}/f_{turb}\approx 1$. 
Finally, we investigate the physical reasons justifying the value of $f_{flap}$ measured along the horizontal branch of the diagram, when $f_{nat}/f_{turb}\ll1$ and the fiber always oscillates at the same frequency. 
We observe that the time scale $1/f_{flap} \approx 2 {H}/{U_b}$ is comparable to the turnover time of the largest eddies in the flow, which span the whole width of the channel and contain most of the overall turbulent kinetic energy.
Simulating an additional turbulent channel flow at $Re=10000$, we confirm that the flapping frequency of a fiber satisfying $f_{nat}/f_{turb}\ll1$ is proportional to the largest flow scale and not to the turbulent eddies of size comparable to the fiber length.
High-$Re$ results are reported as gray squares in Fig.~\ref{fig:freqCollapse}.

The existence of two distinct dynamical regimes for the fiber is in qualitative agreement with previous works studying free fibers suspended in homogeneous isotropic turbulence (HIT) \cite{rosti-etal-2018, olivieri-etal-2020-2, brizzolara-etal-2021, olivieri-mazzino-rosti-2022}, notwithstanding the anisotropic and non-homogeneous nature of the flow we considered here, thus hinting towards a more general behavior of slender flexible bodies in turbulent flows.
Nevertheless, a substantial quantitative difference here is that the fiber is clamped to the wall, which hinders its motion and prevents it from following the flow in a Lagrangian way.
Because of this, the flapping state of the fiber non-trivially relates to the largest scale of the flow and not to the turbulent eddies of comparable size, as previously found for the free fibers \cite{rosti-etal-2018, olivieri-etal-2020-2, brizzolara-etal-2021, olivieri-mazzino-rosti-2022}.
This behavior, contrasting with any previous result, is observed here for the first time and is the main outcome of this Letter.

In this study we have shown that a clamped flexible fiber immersed in a turbulent wall flow can exhibit only two different regimes of motion, depending on how its structural properties (and hence its natural frequency, $f_{nat}$) relate to a characteristic turbulent frequency proportional to the largest scale of the flow, $f_{turb}$.
For $f_{nat}/f_{turb}\gg1$ the fiber oscillates at its natural frequency, while for $f_{nat}/f_{turb}\ll1$ it sways at $f_{turb}$ regardless of its structural parameters.
We also notice that the existence of a state where the turbulent fluctuations are passively followed by the fiber, as highlighted by the PDFs in Fig.~\ref{fig:PDFs}, retains practical relevance, since quantities of engineering interest, like the flow rate or the Reynolds number, can be measured relying on this phenomenon. \\ \\

\begin{acknowledgments}
The research was supported by the Okinawa Institute of Science and Technology Graduate University (OIST) with subsidy funding from the Cabinet Office, Government of Japan. The authors acknowledge the computer time provided by the Scientific Computing section of Research Support Division at OIST, and by RIKEN, under the HPCI System Research Project grant hp220402 on the FUGAKU supercomputer.
SO acknowledges the support by grants FJC2021-047652-I and PID2022-142135NA-I00 by MCIN/AEI/10.13039/501100011033 and European Union NextGenerationEU/PRTR.
RA thanks OIST for the Research Internship program, during which she contributed to this work.
\end{acknowledgments}

\end{document}


\preprint{APS/123-QED}

\begin{center}
	\textit{Supplementary Material for}
\end{center}

\title{Forced and natural dynamics of a clamped flexible fiber in wall turbulence}

\author{Giulio Foggi Rota \orcidA{}}
\author{Morie Koseki \orcidB{}}
\author{Riya Agrawal \orcidC{}}
\altaffiliation[Currently at ]{United College of Engineering and Research, Allahabad, 211010, Uttar Pradesh, India.}
\author{Stefano Olivieri \orcidD{}}
\altaffiliation[Currently at ]{Universidad Carlos III de Madrid, Leganés, 28911, Madrid, Spain.}
\author{Marco Edoardo Rosti \orcidE{}}
\email[E-mail for correspondence: ]{marco.rosti@oist.jp}
\affiliation{Complex Fluids and Flows Unit, Okinawa Institute of Science and Technology Graduate University, 1919-1 Tancha, Onna-son, Okinawa 904-0495, Japan.}

\maketitle

\setcounter{table}{0}
\makeatletter 
\renewcommand{\thetable}{S\@arabic\c@table}
\makeatother

\setcounter{figure}{0}
\makeatletter 
\renewcommand{\thefigure}{S\@arabic\c@figure}
\makeatother

\setcounter{equation}{0}
\makeatletter 
\renewcommand{\theequation}{S\@arabic\c@equation}
\makeatother

\section{Numerical methods}

The motion of an incompressible fluid is governed by the incompressibility constraint and the momentum balance,
\begin{eqnarray}
   &\nabla \cdot \mathbf{u} = 0,
   \label{eq:mass}\\
   &\displaystyle \frac{\partial \mathbf{u}}{\partial t} +\mathbf{u} \cdot \nabla \mathbf{u} = - \frac{1}{\rho_f} \nabla p + \nu \nabla^2 \mathbf{u} + \mathbf{f}_{\mathbf{\rm{for}}} + \mathbf{f}_{\mathbf{\rm{fib}}},
   \label{eq:momentum}
\end{eqnarray}
where $\mathbf{u}(\mathbf{x},t)$ and $p(\mathbf{x},t)$ denote the velocity and pressure fields, $\rho _f$ is the volumetric density and $\nu$ the kinematic viscosity of the fluid. $\mathbf{f}_{\mathbf{\rm{for}}}$ and $\mathbf{f}_{\mathbf{\rm{fib}}}$ represent the forcing term needed to sustain a fully developed turbulent flow and that to account for the fluid-structure interaction, respectively \cite{rosti-etal-2018}. The no-slip and no-penetration boundary conditions are enforced at the walls, while periodicity is imposed in the streamwise and spanwise directions. 
In this study, we tackle equations (\ref{eq:mass},\ref{eq:momentum}) numerically by means of the well tested solver \textit{Fujin} (\href{https://groups.oist.jp/cffu/code}{https://groups.oist.jp/cffu/code}). 
The flow variables are sampled on a staggered Cartesian grid and all space derivatives are discretised with a second order central finite difference scheme; the solution is thus advanced in time with a second order Adams-Bashforth method \cite{mazzino-rosti-2021}.
We resort to a projection-correction algorithm \cite{kim-moin-1985}, solving the Poisson equation for the pressure with an efficient decomposition library (\textit{2decomp}) coupled to an in-place spectral solver based on the Fourier’s series method described by \citet{dorr-1970}.

Let us now consider a flexible fiber of length $h$ and cross-section diameter $d$ clamped at the wall, with homogeneous structural properties.
Its dynamics is described by an extended version of the distributed-Lagrange-multiplier/fictitious-domain (DLM/FD) formulation of the continuum equations \cite{yu-2005}, which represents a generalisation of the Euler-Bernoulli beam model in that it allows for finite deflections, but retains the inextensibility constraint \cite{banaei-rosti-brandt-2020}:
\begin{eqnarray}
    &\Delta \Tilde{\rho} \displaystyle \frac{\partial^2 \mathbf{X}}{\partial t^2} = 
    \displaystyle \frac{\partial}{\partial s}\left(T\displaystyle \frac{\partial \mathbf{X}}{\partial s}\right) - \gamma \displaystyle \frac{\partial^4 \mathbf{X}}{\partial s^4} - \mathbf{F} 
    \label{eq:eulerBernoulli}\\
    &\displaystyle \frac{\partial \mathbf{X}}{\partial s} \cdot \displaystyle \frac{\partial \mathbf{X}}{\partial s} = 1
    \label{eq:inextensibility}
\end{eqnarray}
$\mathbf{X}(s,t)$ is the position of a point on the neutral axis of the fiber as a function of the curvilinear abscissa, $s$, and time, $t$, $ \Delta \Tilde{\rho} = (\rho_s-\rho_f)\pi d^2/4 $ represents the difference among the linear density of the fiber and that of the fluid, $T$ is the tension needed to enforce inextensibility, $\gamma$ is the bending rigidity of the fiber and $\mathbf{F}$ is the fluid-structure coupling term.
We impose $\mathbf{X}\rvert_{s=0}=\mathbf{X_0}$ and ${\partial\mathbf{X}}/{\partial s}\rvert_{s=0}=(0,1,0)$ at the clamp, while enforcing ${\partial^3\mathbf{X}}/{\partial s^3}\rvert_{s=h}={\partial^2\mathbf{X}}/{\partial s^2}\rvert_{s=h}=\mathbf{0}$ and $T\rvert_{s=h}=0$ at the free end. 
Equations (\ref{eq:eulerBernoulli},\ref{eq:inextensibility}) are solved following an approach similar to that of \citet{huang-etal-2007}, but the bending term is treated implicitly to allow for a larger time step \cite{banaei-rosti-brandt-2020}. The resulting linear system is solved with the \textit{dgesv} routine of \textit{LAPACK}.

Finally, we couple the fluid and the structure at their interface through a no-slip boundary condition, $\dot{\mathbf{X}}=\mathbf{u}[\mathbf{X}(s,t),t]$, guaranteed applying the force distribution computed with a Lagrangian immersed boundary method \cite{peskin-2002, huang-etal-2007, banaei-rosti-brandt-2020}.
Our solver is the same previously employed by \citet{monti-olivieri-rosti-2023}, who validated the motion of a single flexible fiber against the analytical predictions reported by \citet{huang-etal-2007} and the mean flow quantities above and between a rigid canopy against the experiments of \citet{yoshihiko-etal-1991}.

\pagebreak

\section{Setup}

\begin{figure}[H]
\centering
\begin{minipage}{0.3\textwidth}
\vspace*{-1cm}
\hspace*{.70cm}
\includegraphics[scale=0.3]{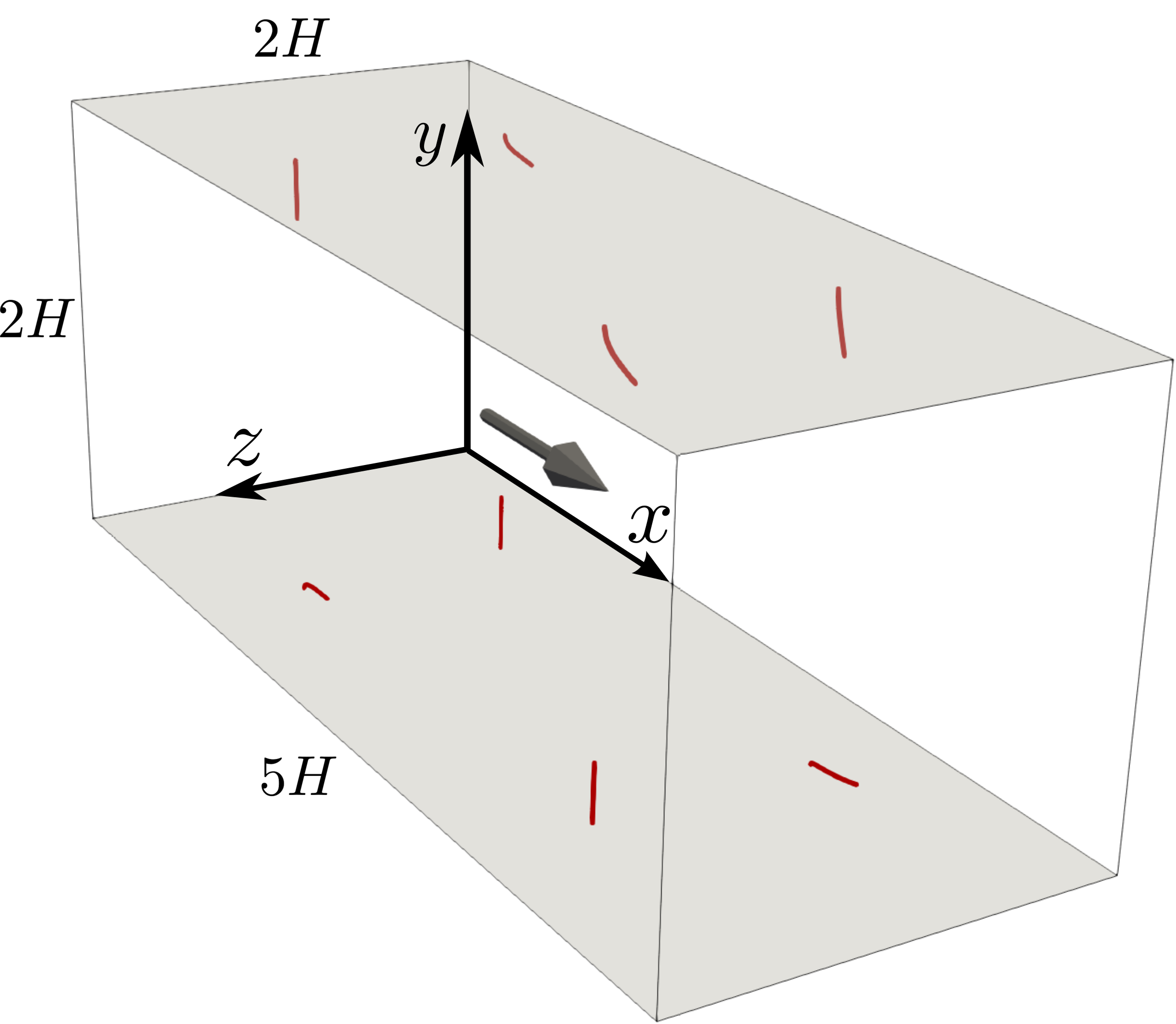}
\put(-180,130){a)}
\end{minipage}\hfill
\begin{minipage}{0.45\textwidth}
\includegraphics[scale=0.55]{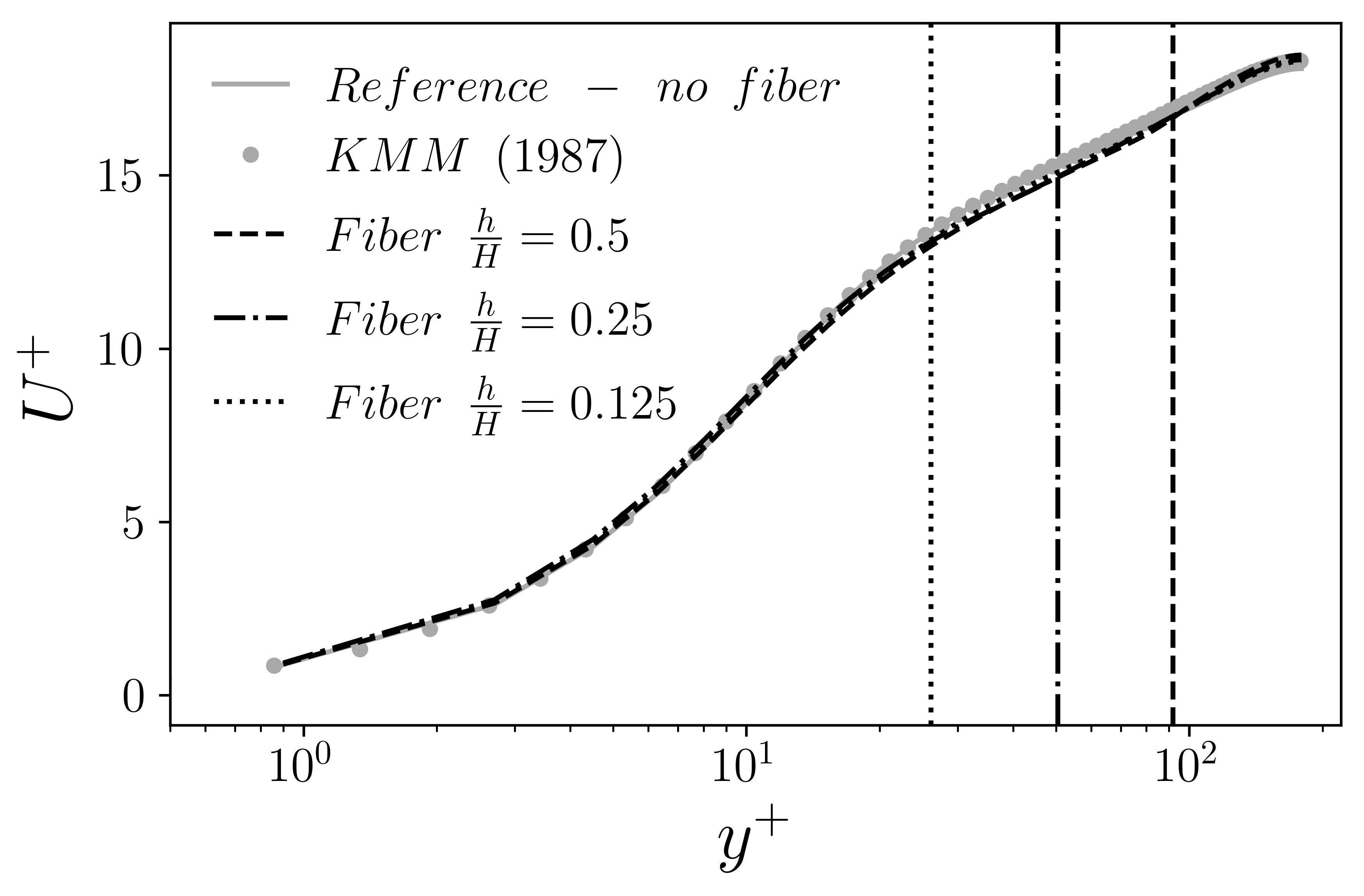}
\put(-240,160){b)}
\end{minipage}\hfill
\caption{\justifying Panel a reports a sketch of the computational domain adopted throughout our simulations, populated with the fibers of case 3 in table \ref{tab:parameters}. A gray arrow denotes the flow direction. In panel b we show the mean velocity profiles in wall units from our simulations, interpolated at the centre of each grid cell (case 2 from table \ref{tab:parameters} is omitted, since it yields an outcome which is identical to that of case 1); a reference simulation without fibers is also included. Results are compared to the data of \citet{kim-moin-moser-1987} in order to asses the adequacy of our numerical setup. Vertical lines mark the vertical coordinate corresponding to the fiber length $h$ of the different cases.}
\label{fig:validation}
\end{figure}

Our simulations are carried out in a rectangular volume bounded from above and below by two flat walls extending along the streamwise ($x$) and spanwise ($z$) directions.
Starting from a domain of size $2.5H \times 2H \times H$ (in $x$, $y$ and $z$, respectively) homogeneously discretised over $192 \times  160 \times 96$ points, we later moved to a larger one (Fig. \ref{fig:validation}a) of size $5H \times 2H \times 2H$ for which we considered two homogeneous discretisations made of $500 \times 200 \times 200$ and $1000 \times 400 \times 400$ points.
All grids yielded consistent results for the dynamics of most fibres, but for the few most rigid ones. Given the long computational times needed for the convergence of the statistics associated to the most flexible fibres, we therefore opted for the intermediate resolution introduced above, complementing the outcome with shorter high-resolution simulations for the most rigid fibres.
Imposing a fixed flow rate so that  $Re=U_b 2H/\nu=5600$, we confirm that the size of our domain is larger than the minimum one needed to sustain turbulence in numerical simulations \cite{jimenez-moin-1991}.
After testing both lower and higher values, we ran all of our simulations at $\Delta t=1e^{-5} H/U_b$. 
This time-step proves appropriate to correctly resolve the fastest dynamics of the fibers and of the flow, resulting in a Courant number of $\mathcal{O}(1e^{-3})$ on the finest grid. 

The channel is populated with flexible fibers of different rigidities, vertically clamped at the upper and lower walls and protruding towards the centreline (Fig. \ref{fig:validation}a). 
We consider logarithmically distributed values of the $Ca$ down to a minimum dictated by the capability of our code to resolve the motion of the most rigid fibers. 
Simulations are repeated for different fiber lengths ($h$) and density ratios between fiber and fluid ($\rho_s/\rho_f$), thus spanning the parameter space reported in table \ref{tab:parameters}.
After assessing the independence of the fibre motion from the number of Lagrangian points, we choose it as in table \ref{tab:parameters}, to have a Lagrangian grid spacing comparable to the Eulerian one.

Since the purpose of our study is to investigate the dynamics of a single flexible fiber in wall turbulence, rather than that of an hairy surface, we must ensure that the fibers contained in the computational domain are sufficiently far from each other not to affect the mean flow significantly, and therefore behave as independent entities.
We verify this in Fig. \ref{fig:validation}b, by comparing the mean velocity profiles from our simulations to those reported by \citet{kim-moin-moser-1987} for a turbulent channel flow at the same Reynolds number.
All the curves shown in the figure exhibit a good agreement, thus validating our numerical setup.

\begin{table}[H]
    \centering
    \caption{Parameter space investigated with our simulations.}
    \setlength{\tabcolsep}{12pt}
    \begin{tabular}{ccccc}
    \hline \hline
       Case & $h/H$ & $\rho_s/\rho_f$ & $Ca_{min}\backslash Ca_{max}$ & \# Lagrangian points \\ \hline
       1 & 0.5      & $1.08$ & $2.5 \cdot 10^{-2}\backslash 2.5 \cdot 10^{5}$ & 40\\
       2 & 0.5      & $1.8$ & $2.5 \cdot 10^{-2}\backslash 2.5 \cdot 10^{4}$ & 40\\
       3 & 0.25    & $1.08$ & $3.1 \cdot 10^{-3}\backslash 3.1 \cdot 10^{3}$ & 20\\
       4 & 0.125  & $1.08$ & $3.9 \cdot 10^{-4}\backslash 3.9 \cdot 10^{-1}$ & 10\\ 
    \hline \hline
    \end{tabular}
    \label{tab:parameters}
\end{table}

\pagebreak

\section{Time histories}

\begin{figure}[h!]
\centering
\includegraphics[scale=0.43]{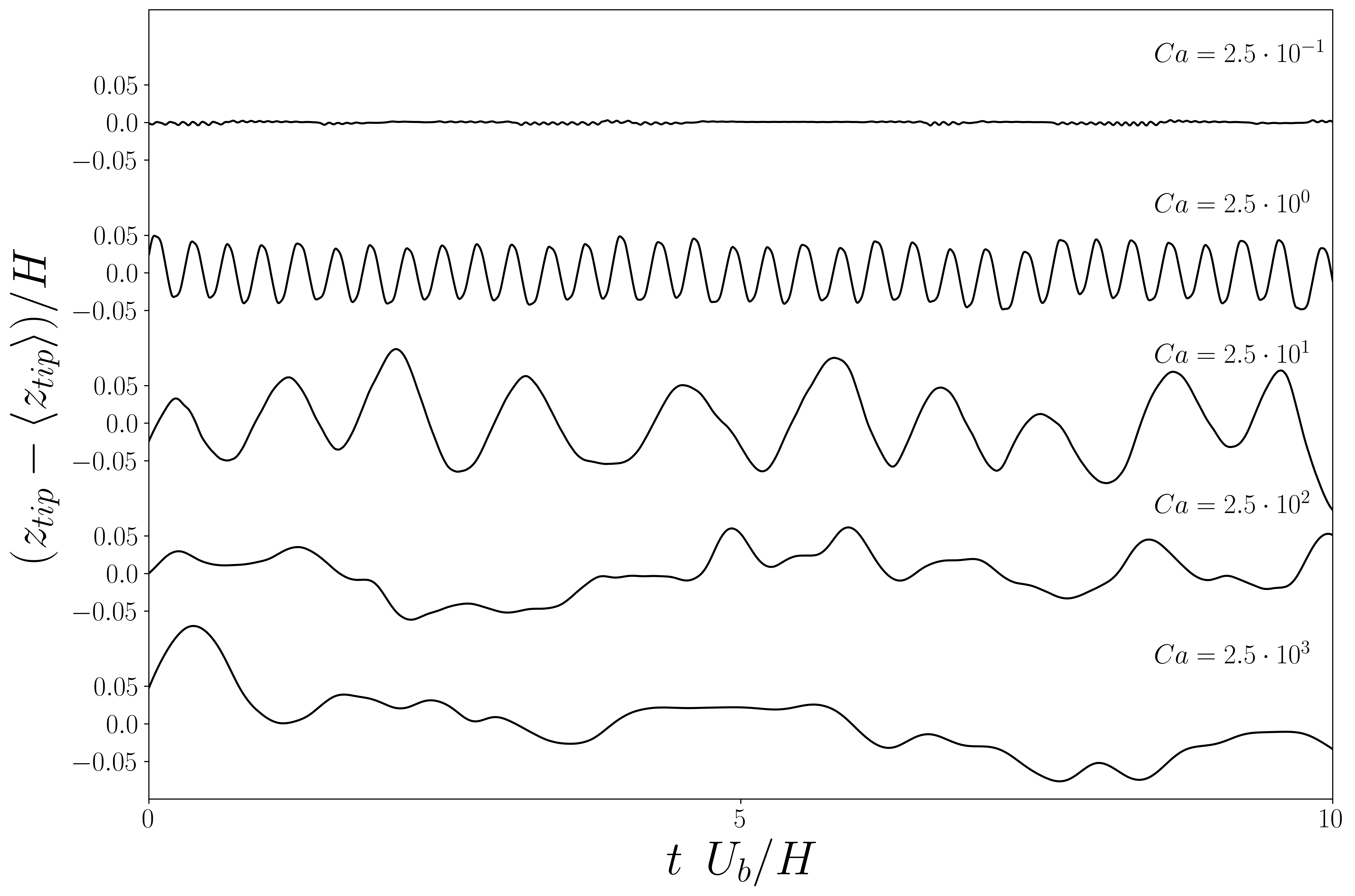}
\caption{\justifying Spanwise position of the tip of five fibers from case 1 of table \ref{tab:parameters}, for increasing values of $Ca$. The lateral oscillation amplitude of the fiber with $Ca=2.5\cdot 10^{1}$ (for which $f_{flap}\approx f_{nat}\approx f_{turb}$) appears larger than that of its neighbors, suggesting a resonance effect.}
\label{fig:tHist}
\end{figure}

Here we observe the time evolution of the spanwise tip position for five fibers of decreasing rigidity, two laying in the $f_{flap}\approx f_{nat}$ regime, one in the $f_{flap}\approx f_{nat}\approx f_{turb}$ regime, and two in the $f_{flap}\approx f_{turb}$ regime of the map in Fig. 6. 
Only ten bulk time instants are considered for the purpose of this plot, even thought our simulations were run for much longer: we sampled the fiber position and velocity every $\tilde{\Delta t}=1e^{-3} H/U_b$ over a total time of $300 H/U_b$ for the more flexible fibers (characterised by a slower dynamics) and $100 U_b/H$ for the more rigid ones (characterised by a faster dynamics), excluding the initial transients.
The two more rigid fibers in Fig. \ref{fig:tHist} oscillate at their natural frequencies with a nearly-sinusoidal motion, consistently with the bimodal $PFD$ in Fig. 3b. 
The two more flexible ones, instead, sway with a time scale of $\mathcal{O}(H/U_b)$. 
In general, the lateral oscillation amplitude of the fibers grows for increasing values of $Ca$ (Fig. 2b). 
Nevertheless, we also observe a further moderate increase in the intermediate regime where $f_{flap}\approx f_{nat}\approx f_{turb}$, suggesting a resonance effect.
This behaviour appears consistent with the elastic energy peak noticed by \citet{rosti-etal-2018} for intermediate values of the rigidity.

\section{High-frequency scaling of the temporal spectra}

\begin{figure}
\centering
\includegraphics[scale=0.7]{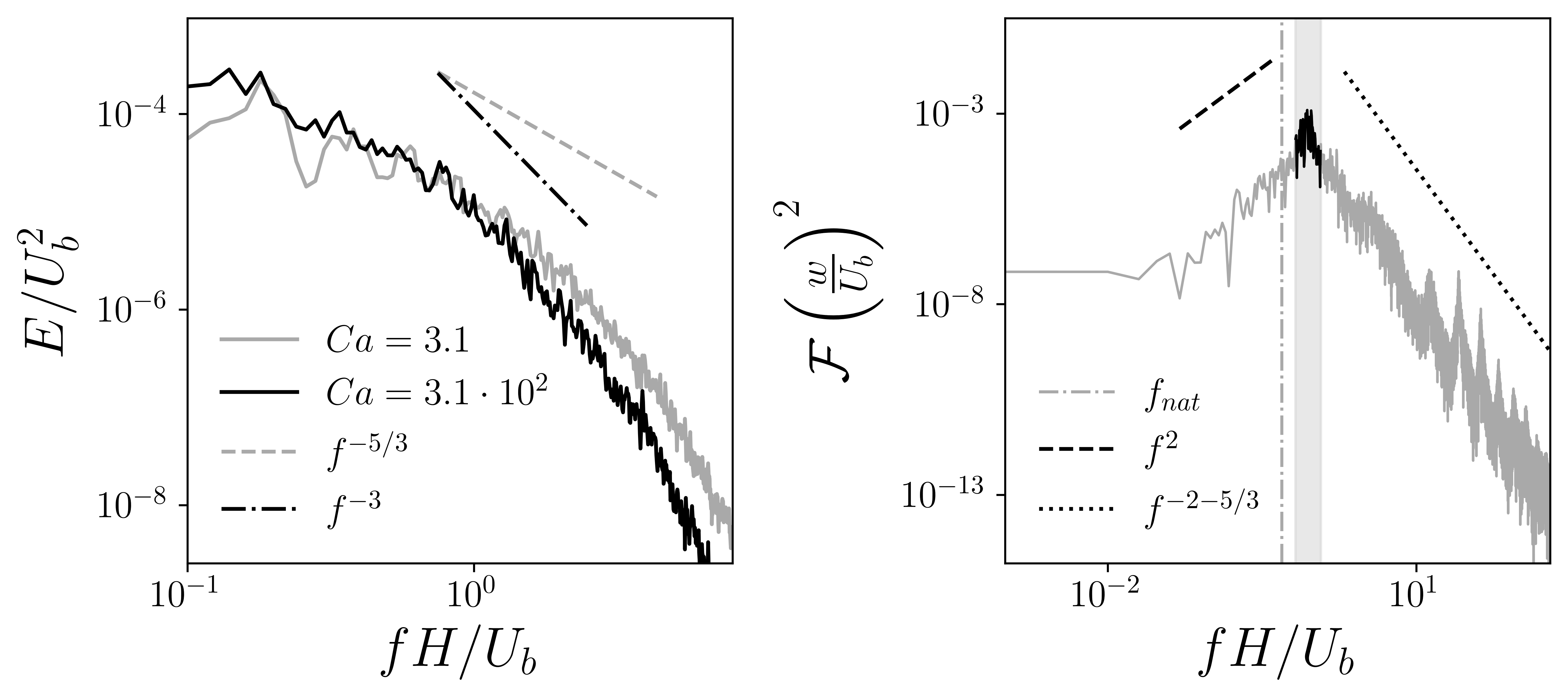}
\put(-430,178){a)}
\put(-218,178){b)}
\caption{\justifying Further details on the high-frequency scaling of the temporal spectra. In panel $a$ we report the Eulerian temporal spectra of the turbulent kinetic energy ($E$) at a wall distance corresponding to the mean deflected tip positions of a more rigid  ($Ca=3.1$, $h=0.25H$, $\rho_s=1.08\rho_f$) and a more flexible ($Ca=3.1\cdot 10^{2}$, $h=0.25H$, $\rho_s=1.08\rho_f$) fiber. The former tends towards $f^{-5/3}$, while the latter exhibits a sharper decay compatible with $f^{-3}$. Panel $b$ shows the Lagrangian temporal spectrum of the fiber tip spanwise velocity for a fiber ($Ca=25$, $h=0.5H$, $\rho_s=1.08\rho_f$) clamped vertically at $y/H=0.075$, in the middle of the flow. While the fiber is oscillating at the turbulent frequency dictated by the flow, a $\xi\approx-5/3$ decay is observed, thus confirming the independence of the scaling from the regime of oscillation of the fiber.}
\label{fig:valSpectra}
\end{figure}

After reaching their respective maxima at a frequency close to $f_{turb}$ or $f_{nat}$, respectively, the temporal spectra of the spanwise fiber tip velocity shown in Fig. 4 of the main text are characterised by a decay $f^{-2+\xi}$ that we claim to depend on the forcing seen by the fiber in the flow region it spans during its motion, consistently with the arguments of \citet{jin-ji-chamorro-2016}. 
In particular, the more flexible fiber is deflected in the  lower part of the viscous wall region and it is forced by a smooth turbulent field, determining the $\xi\approx-3$ scaling of the Lagrangian spectrum, while the more rigid one extends in the buffer layer, recovering $\xi\approx-5/3$. 
To support these statements, we compute the temporal spectra of the turbulent kinetic energy ($E$) at the wall distances corresponding to the mean positions of the deflected fiber tips.
As visible in panel $a$ of Fig. \ref{fig:valSpectra}, the Eulerian spectrum at the position corresponding to the fiber with $Ca=3.1$ tends towards $f^{-5/3}$, while that at $Ca=3.1\cdot 10^{2}$ exhibits a steeper decay.
We impute the sharper trends observed in the case of the Lagrangian spectra to the integral effect exerted by the fiber on the high frequency fluctuations.

Furthermore, we prove that the slope of the high frequency region of the spectra is independent from the regime of motion of the fiber. 
By clamping vertically one of the fibers with intermediate flexibility of case 1 (see table \ref{tab:parameters}) in the middle of the flow, at a distance from the wall of $y/H=0.075$, its tip lays well within the buffer layer.
A $\xi\approx-5/3$ decay is observed in this case for the temporal spectrum of the spanwise fiber tip velocity, even if the fiber is swaying at the turbulent frequency dictated by the flow and not at its natural one, $f_{nat}$.
The Lagrangian spectrum reported in panel $b$ of Fig. \ref{fig:valSpectra} therefore confirms that the high frequency region is dominated by the forcing to which the fiber is exposed.

\bibliography{./../Wallturb.bib}